\documentclass[twocolumn,showpacs,preprintnumbers,amsmath,amssymb,showkeys,superscriptaddress]{revtex4}

\usepackage{graphicx}
\usepackage{dcolumn}
\usepackage{bm}
\usepackage{longtable}

\begin{document}

\preprint{APS/123-QED}

\title{Measurement of the radiative neutron capture cross section of $^{206}$Pb and its
  astrophysical implications}

\author{C.~Domingo-Pardo} \thanks{Corresponding author. E-mail: cesar.domingo.pardo@cern.ch}  \affiliation{Forschungszentrum Karlsruhe GmbH (FZK), Institut f\"{u}r Kernphysik, Germany} \affiliation{Instituto de F{\'{\i}}sica Corpuscular, CSIC-Universidad de Valencia, Spain}   
\author{U.~Abbondanno}  \affiliation{Istituto Nazionale di Fisica Nucleare (INFN), Trieste, Italy}   
\author{G.~Aerts}  \affiliation{CEA/Saclay - DSM/DAPNIA, Gif-sur-Yvette, France}   
\author{H.~\'{A}lvarez}  \affiliation{Universidade de Santiago de Compostela, Spain}   
\author{F.~Alvarez-Velarde}  \affiliation{Centro de Investigaciones Energeticas Medioambientales y Technologicas, Madrid, Spain}   
\author{S.~Andriamonje}  \affiliation{CEA/Saclay - DSM/DAPNIA, Gif-sur-Yvette, France}   
\author{J.~Andrzejewski}  \affiliation{University of Lodz, Lodz, Poland}   
\author{P.~Assimakopoulos }  \affiliation{University of Ioannina, Greece}   
\author{L.~Audouin }  \affiliation{Forschungszentrum Karlsruhe GmbH (FZK), Institut f\"{u}r Kernphysik, Germany}   
\author{G.~Badurek}  \affiliation{Atominstitut der \"{O}sterreichischen Universit\"{a}ten,Technische Universit\"{a}t Wien, Austria}   
% ...

\collaboration{The n\_TOF Collaboration} 

\date{\today}%

\begin{abstract}

The ($n, \gamma)$ cross section of $^{206}$Pb has been measured at the
CERN n\_TOF facility with high resolution in the energy range from
1~eV to 620~keV by using two optimized C$_6$D$_6$ detectors. In the
investigated energy interval about 130 resonances could be observed, from which 
61 had enough statistics to be reliably analyzed via the R-matrix analysis
code SAMMY. Experimental  uncertainties were minimized, in particular with
respect to (i) angular  distribution effects of the prompt capture
$\gamma$-rays, and to (ii)  the TOF-dependent background due to
sample-scattered neutrons. 
Other background components were addressed by background measurements 
with an enriched $^{208}$Pb sample. The effect of the lower energy 
cutoff in the pulse height spectra of the C$_6$D$_6$ detectors was 
carefully corrected via Monte Carlo simulations. Compared to
previous $^{206}$Pb values, the Maxwellian averaged capture cross 
sections derived from these data are about 20\% and 9\% lower at 
thermal energies of 5~keV and 30~keV, respectively. These new results 
have a direct impact on the $s$-process abundance of $^{206}$Pb, which 
represents an important test for the interpretation of the cosmic
clock based on the decay of $^{238}$U.

\end{abstract}

\maketitle

\section{\label{sec:introduction} Introduction}
Similar to the majority of the stable isotopes beyond iron, 
$^{206,207,208}$Pb and $^{209}$Bi are synthesized by the rapid
\mbox{($r$-)} and slow \mbox{($s$-)} neutron capture processes. 
However, this mass region is particularly interesting because the
$r$-process abundances are dominated by the decay of
the short lived $\alpha$-unstable transbismuth isotopes
\cite{cow99}. 
This feature provides an important consistency check for the 
$r$-process abundance calculations in the actinide region,
since the integrated $r$ residuals are constrained by the 
difference between the solar abundance values and the 
respective $s$-process components. Reliable $r$-process 
calculations are required for the interpretation of the
observed Th and U abundances in the ultra metal-poor (UMP)
stars of the Galactic halo. Since these stars are considered 
to be as old as the Galaxy, the observed Th and U abundances 
can be used as cosmo-chronometers, provided the original 
Th and U abundances are inferred from $r$-process models. 
This dating mechanism has the advantage of being independent 
of the yet uncertain $r$-process site~\cite{cow99,sch02,kra04}.

Apart from its relevance for establishing the basic constraints 
for the $r$-process chronometry in general, $^{206}$Pb contains 
also dating information in itself. The $^{206}$Pb/$^{238}$U 
cosmochronometer was first introduced by Clayton in 1964~\cite{cla64}.
The $^{238}$U produced by the $r$ process decays with a half life 
of $t_{1/2}=4.5\times 10^9$ yr over a chain of $\alpha$ and 
$\beta$ decays ending at $^{206}$Pb. Therefore, its radiogenic abundance 
component, $N^{206}_{c}$, can be used to constrain 
the age of the parent isotope $^{238}$U, and hence the age ($\Delta_r$) of 
the $r$-process. Unlike the more direct $r$-process abundance 
predictions derived from the Th and U abundances in UMP stars,
this procedure requires a Galactic evolution model, which 
describes the supernova rate or the frequency of the $r$-process 
events~\cite{fow60}. The drawback of this clock 
arises from the difficulty 
to isolate the cosmo-radiogenic component of $^{206}$Pb accurately 
enough from the additional abundance components. 

Apart from these astrophysical aspects, the neutron capture cross 
section of $^{206}$Pb is also of importance for the design of fast 
reactor systems based on a Pb/Bi spallation source. Because
24.1\% of natural lead consists of $^{206}$Pb, its ($n, \gamma$)
cross section influences the neutron balance of the reactor~\cite{ado03}.

There have been several measurements of the $^{206}$Pb($n, \gamma$) cross section, which show discrepancies that are
difficult to understand (see Sec.~\ref{sec:comp}). The aim of this work is to
perform a new independent measurement with higher accuracy and in this way 
to determine the $s$-process contribution to the $^{206}$Pb
abundance, $N^{206}_s$, more reliably.

In fact, the $s$-process abundance of this isotope is almost
completely determined by the stellar ($n, \gamma$) cross section,
nearly independent of the stellar model used \cite{rat04}. Therefore, 
the uncertainty of $N^{206}_s$ arises mostly from the cross 
section uncertainty.



Potential sources of systematic error have been substantially reduced in
the present measurement, which was performed at the CERN n\_TOF
installation. The new setup, and in particular the detectors themselves, were
optimized for very low neutron sensitivity. Furthermore, the detectors were 
mounted at $\sim$125$^\circ$ with respect to the incident neutron 
beam in order to minimize the correction for angular distribution 
effects. The experimental details are presented in 
Sec.~\ref{sec:measurement}, followed by the adopted data analysis 
procedures and an evaluation of the various systematic uncertainties
in Sec.~\ref{sec:analysis}. The deduced resonance parameters and the 
corresponding Maxwellian averaged capture cross sections in the 
stellar temperature regime are presented in Sec.~\ref{sec:results}. 
Based on these new data, first astrophysical implications for the 
$s$-process abundance of $^{206}$Pb are discussed in 
Sec.~\ref{sec:implications}. 

\section{\label{sec:measurement} Measurement}
The time-of-flight (TOF) measurement was performed at the CERN n\_TOF
installation~\cite{preport} using a set of two C$_6$D$_6$ detectors.
Neutrons were produced by a 20~GeV proton beam on a lead spallation 
target. The spallation source was surrounded by a 6~cm thick water 
layer, which served as a coolant and as a moderator for the initially 
fast neutron spectrum. The beam was characterized by intense bunches 
of (3 to 7)$\times$10$^{12}$ protons, a width of 6~ns (rms), and
a repetition rate of only 0.4 Hz. This extremely low duty-cycle allows 
one to perform ($n, \gamma$) measurements over a broad neutron energy 
interval from 1~eV up to 1~MeV and to achieve favorable background 
conditions. Data were recorded by means of an advanced acquisition 
system with zero dead time, based on 8~bit Flash-Analog-to-Digital 
Converters (FADC), with 500~MHz sampling rate and 8~MB buffer 
memory~\cite{abb05}.

The measurement was performed with an enriched metal sample 8.123~g
in mass and 20~mm in diameter. The sample was enriched to 99.76\% in 
$^{206}$Pb with small impurities of $^{207}$Pb (0.21\%) and 
$^{208}$Pb (0.03\%).

Capture events were registered with two C$_6$D$_6$ $\gamma$-ray detectors 
optimized for very low neutron sensitivity~\cite{pla03}. A sketch 
of the experimental setup is shown in Fig.~2 of Ref.~\cite{dom06}. 
The absolute value of the neutron fluence was determined by regular 
calibration measurements with an 0.5~mm thick gold sample and by 
using the saturated resonance technique~\cite{mac79} for the first 
gold resonance at $E_n=4.9$~eV. The energy differential neutron flux 
was determined with a relative uncertainty of $\pm$2\% from the
flux measurement with a $^{235,238}$U fission chamber calibrated by 
Physikalisch-Technische Bundesanstalt (PTB)~\cite{ptb}. The
neutron intensity at the sample position was also monitored by means
of a \mbox{200-$\mu$g/cm$^2$} thick $^6$Li foil in the neutron beam
about 2.5~m upstream of the capture sample. The $^6$Li foil was
surrounded by four silicon detectors outside of the beam for recording
the $^3$H and $\alpha$ particles from the ($n, \alpha$) reactions.

Compared to previous measurements~\cite{miz79,all73}, the present setup
had the advantage that the detectors were placed at $\sim$125$^{\circ}$ 
with respect to the incident neutron beam. In this way, the corrections
for angular distribution effects of the prompt capture $\gamma$-rays 
were strongly reduced. This configuration led also to a substantial 
reduction of the background from in-beam $\gamma$-rays scattered 
in the sample~\cite{abb03}.

\section{\label{sec:analysis} Capture data analysis}
The response function of the C$_6$D$_6$ detectors needs to be modified 
such that the detection probability for capture cascades becomes independent 
of the cascade multipolarity. This was accomplished by application 
of the Pulse Height Weighting Technique (PHWT)~\cite{mac67}. Based on 
previous experience~\cite{abb04,dom06,dom06b}, the weighting functions 
(WFs) for the gold and lead samples were obtained by means of Monte Carlo
calculations. The accuracy of the WFs was verified with the method
described in Ref.~\cite{abb04}, by which the calculated WFs were applied
to Monte Carlo simulated capture $\gamma$-ray spectra. Using this 
procedure, the uncertainty of the WFs was estimated to be smaller than 
0.5\% for the samples used in the present experiment. 


The weighted count rate $N^w$ is then transformed into an experimental 
yield,

\begin{equation}\label{eq:yexp}
Y^{exp} = f^{sat} \frac{N^w}{N_n E_c},
\end{equation}

\noindent
where the yield-normalization factor $f^{sat}$ is determined by
calibration measurements using the saturated 4.9~eV resonance in gold. 
$N_n$ denotes the neutron flux and $E_c$ the effective binding energy.

The yield in Eq.~(\ref{eq:yexp}) is still subject to several corrections. 
The common effects of the background and of the low energy cutoff in 
the pulse height spectra of the $\gamma$ detectors are described in 
Secs.~\ref{sec:background} and \ref{sec:threshold}, respectively. 
The measurement on $^{206}$Pb is particularly sensitive to the 
angular distribution of the prompt capture $\gamma$-rays. The 
impact of this effect is described in Sec.~\ref{sec:angular}.

\subsection{\label{sec:background}Backgrounds}
A major source of background is due to in-beam $\gamma$-rays, 
predominantly from neutron captures in the water moderator, which 
travel along the neutron flight tube and are scattered in the 
$^{206}$Pb sample. This background exhibits
a smooth dependence on neutron energy, with a broad maximum around 
$E_n \approx 10$~keV. The shape of this background was determined 
from the spectrum measured with an isotopically pure $^{208}$Pb sample, 
which contains practically no resonances in the investigated neutron 
energy range. This spectrum was properly scaled and used as a point-wise 
numerical function in the R-matrix analysis of the $^{206}$Pb capture 
yield (see Sec.~\ref{sec:results}).

Another type of background arises in the analysis of resonances with a
dominant neutron scattering channel, $\Gamma_n >> \Gamma_\gamma$. In 
such cases, there are about $\Gamma_n/\Gamma_\gamma$ scattered neutrons 
per capture event. These scattered neutrons can be captured in the 
detectors or in surrounding materials, thus mimicking true capture 
events. This effect was estimated to be negligible for all the resonances
listed in Table~\ref{tab:RP}. 

\subsection{\label{sec:threshold}Digital threshold}

As mentioned in Sec.~\ref{sec:measurement}, FADCs were used for 
recording directly the analog output signals of the C$_6$D$_6$
detectors. Without any further discrimination, 8~MB of data
would have been acquired per proton pulse in each detector. 
Depending on the sample, this enormous amount of data could be 
reduced by factors of 20 to 100 by using a zero suppression 
algorithm (see Ref.~\cite{abb05} for details). By this method 
events below a certain pulse-height are discriminated by a 
constant digital threshold analogous to conventional data 
acquisition systems, where an electronic threshold is used to
reduce backgrounds and dead time effects.

\begin{figure}[h]
\includegraphics[width=0.45\textwidth]{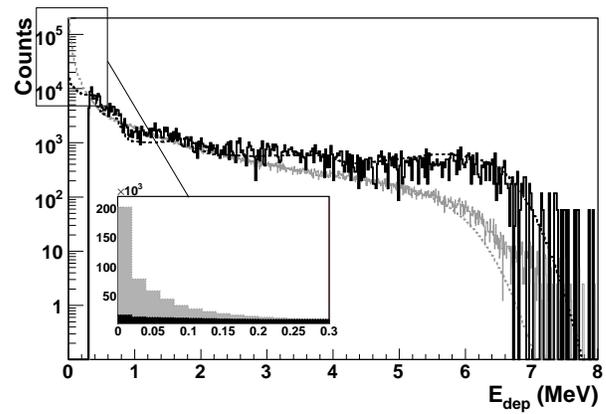}
\caption{\label{fig:PHS} Pulse height spectra for the 4.9~eV resonance in
  gold (grey) and for the 3.3~keV resonance in $^{206}$Pb (black), arbitrarily
  scaled. The dashed lines are the MC-calculated $\gamma$-ray spectra for the 
  two resonances. The linear scale used in the inset illustrates the large 
  difference between the simulated spectra below a threshold of 300~keV.}
\end{figure}

Due to this threshold, the pulse height spectra of the C$_6$D$_6$
detectors exhibit a low energy cutoff at a certain value of the 
signal amplitude (see Fig.~\ref{fig:PHS}). 
In this experiment the threshold was set at a $\gamma$-ray energy
of 320~keV. If the pulse height spectra of the $^{206}$Pb
sample and of the gold sample used for normalization would have 
the same shape, the fraction of weighted counts below this threshold 
would nearly cancel out in the expression for the yield,

\begin{eqnarray}
Y^{exp} \propto \frac{\sum_{0keV}^{320~keV} W^{\rm Pb}_i R^{\rm Pb}_i +
  \sum_{320~keV}^{E_c} W^{\rm Pb}_i R^{\rm Pb}_i}{\sum_{0keV}^{320~keV} W^{\rm Au}_i R^{\rm Au}_i +
  \sum_{320~keV}^{E_c} W^{\rm Au}_i R^{\rm Au}_i} \nonumber \\
\approx \frac{\sum_{320~keV}^{E_c} W^{\rm Pb}_i R^{\rm Pb}_i}{\sum_{320~keV}^{E_c} W^{\rm Au}_i R^{\rm Au}_i}.
\end{eqnarray}

\noindent
Here, the $W_i$ and $R_i$ are the corresponding weighting factors 
and response functions for a certain time of flight channel, 
respectively. However, this approximation is only valid within 4 to 
5\%, because the pulse height spectra of captures 
on $^{206}$Pb and $^{197}$Au  differ significantly near threshold 
(Fig.~\ref{fig:PHS}). 

This effect has been taken into account in the determination of the 
experimental capture yield by simulating the capture cascades of each 
isotope as described in detail in Refs.~\cite{abb04,dom06,dom06b}. 
Fig.~\ref{fig:PHS} shows that the experimental spectra above the 
digital threshold are well reproduced by the simulations. With this 
correction the experimental yield becomes 

\begin{eqnarray}
Y^{exp} \propto \frac{f^t_{\rm Pb}}{f^t_{\rm Au}} \frac{\sum_{320~keV}^{E_c} 
   W^{\rm Pb}_i R^{\rm Pb}_i}{\sum_{320~keV}^{E_c} W^{\rm Au}_i R^{\rm Au}_i}.
\end{eqnarray}

For the adopted digital threshold the yield of the 4.9~eV
resonance in $^{197}$Au needs to be scaled by a factor $f^t_{\rm Au} =
1.071(3)$, whereas the yield of the resonances in $^{206}$Pb required
a correction of $f^t_{\rm Pb} = 1.021(5)$ due to their harder spectrum.
Hence, the correction factor of the final experimental yield was $f^t =
f^t_{\rm Pb}/f^t_{\rm Au} = 0.952(4)$.

\subsection{\label{sec:angular}Angular distribution effects}
Neutron capture with orbital angular momentum $l>0$ leads to an 
aligned state in the compound nucleus, perpendicular to the 
direction of the incident neutron. Given the small multiplicity 
($m=1$ to 2) of the capture cascades in $^{206}$Pb, most of the 
prompt $\gamma$-rays registered with the C$_6$D$_6$ detectors 
still carry this anisotropy, which affects the measured yield.
The angular distribution is in general given by,

\begin{eqnarray}\label{eq:W}
W(\theta) = \sum_k A_k P_k(cos\theta) = 1 + A_2 
  P_2(cos\theta) + \nonumber\\
  + A_4 P_4(cos\theta) + A_6 P_6(cos\theta),
\end{eqnarray}

\noindent
where $P_{k}(cos\theta)$ are the Legendre polynomials of order $k$ 
and $A_{k}$ are coefficients, which depend on the initial ($J$) 
and final ($J'$) spin values, on the multipolarity ($l$) of the 
transition, and on the degree of alignment. The angular distribution 
effects in the capture yield are minimized (although not avoided) 
by setting the detectors at 125$^{\circ}$. Since each C$_6$D$_6$
detector covers a substantial solid angle, capture $\gamma$-rays
are registered around 125$^{\circ} \pm \Delta\theta$. For the 
actual setup of the present measurement one finds
$\Delta\theta \approx 28^{\circ}$. 

\subsubsection{Resonances with spin $J=1/2$}

For resonances with $J = 1/2$ it can be assumed that they 
decay directly to the ground state ($J^{\pi} = 1/2^-$) or to the first or 
second excited states with $J^{\pi} = 5/2^-$ and $J^{\pi} = 3/2^-$, 
respectively (see also Fig.~\ref{fig:levels}). In these cases, one 
finds that $A_2 = A_4 = A_6 = 0$. Therefore, only resonances with 
spin $J > 1/2$ may be affected by angular distribution effects. 


\subsubsection{Resonances with spin $J=3/2$}

In order to quantify the uncertainty due to the angular distribution 
of the prompt $\gamma$-rays emitted from excited states with $J^{\pi} 
= 3/2^-$ the de-excitation patterns reported in Ref.~\cite{miz79} have
been used (Table~\ref{tab:ftheta}). 

\begin{table}
\caption{\label{tab:ftheta} Measured decay patterns from resonances 
with spin $J=3/2$ \cite{miz79}. The systematic uncertainty in the yield 
of each resonance due to the angular distribution of the involved 
transitions are given in the last column.}
\begin{ruledtabular}
\begin{tabular}{lccccc}
$E_{\circ}$ (keV)     & \multicolumn{4}{c}{Intensity $I_{\gamma}$ (\%)} 
                      & $\sigma^{3/2^-}_\theta$\\
\hline
& \multicolumn{4}{c}{$E_{\gamma}$ (keV)}  \\
           &  6737.9  & 6168.6  & 5840.8    & 4114.5   &           \\
\hline
3.36       & 76.0(27) &  2.5(8) &  8.58(11) &  13.0(8) & $\pm$10\% \\
3.36$^a$   &   60     &  2.5    &  24.5     &   13     & $\pm$8\%  \\
10.86      &          &  100    &           &          & $\pm$2\%  \\
21.87      &          &  100    &           &          & $\pm$2\%  \\
42.07      &          &         &  100      &          & $\pm$10\% \\
\end{tabular}
\end{ruledtabular}
$^a$ This work. 
\end{table}

\begin{figure}[h]
\includegraphics[width=0.25\textwidth]{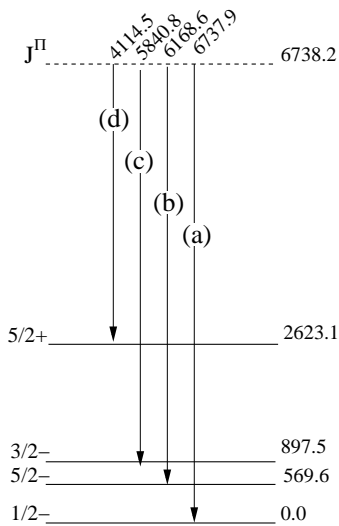}
\caption{\label{fig:levels} Level scheme and decay patterns for $^{207}$Pb
          \cite{miz79}. All energies are in keV.}
\end{figure}

For the first resonance at 3.36~keV, fair agreement has been found
between the relative intensities of Ref.~\cite{miz79} and the rather
coarse values deduced from the experimental pulse height spectrum 
(Table~\ref{tab:ftheta} and Fig.~\ref{fig:PHS}), which suffer from
uncertainties due to background subtraction, limited counting
statistics and poor energy resolution of the C$_6$D$_6$ detectors.
Therefore, an uncertainty of about 20\% has to be ascribed to the
quoted $\gamma$-ray intensities. 

The estimated effect of the angular distribution on the capture 
yield ($\sigma^{3/2^-}_\theta$) is given in the last column of 
Table~\ref{tab:ftheta}. These values were obtained via Monte Carlo 
simulations of the experimental setup, using the energies and 
intensities listed in Table~\ref{tab:ftheta} and the prescription 
of Ref.~\cite{fer65}. The main uncertainty in the calculation of 
the angular distribution effects arises from the unknown admixtures 
of different multipolarities (M1+E2) for the transitions connecting 
the original excited state $J^{\pi} =3/2^-$ with any of the three 
lowest states (paths (a), (b) and (c) in Fig.~\ref{fig:levels}). 
As shown in Table~\ref{tab:ftheta}, the decay pattern and the
corresponding effect on the capture yield $\sigma^{3/2^-}_\theta$ 
vary abruptly from one resonance to another. It is therefore
difficult to assess a common systematic uncertainty for the
remaining ${3/2^-}$ resonances. Assuming that the four resonances 
listed in Table~\ref{tab:ftheta} constitute a representative 
sample, one may consider their standard deviation of $\sigma = 
4$\% as a realistic estimate of the systematic uncertainty due to 
angular distribution effects.

Resonances with $J^{\pi} =3/2^+$ can be assumed to decay directly to the
ground state through an E1 transition. In this case we have estimated an
effect of 10\% in the capture yield with respect to the isotropic case.
However, since $3/2^+$ resonances appear at a relatively high neutron energy,
the final effect in the MACS is practically negligible (see below Sec.~\ref{sec:macs}).

\subsubsection{Resonances with spin $J=5/2$}
For resonances in $^{207}$Pb with $J^{\pi}=5/2^+$ the most probable decay
would be through an electric dipole transition to the first excited state
with $J^{\pi}=5/2^-$ and/or to the second excited state with $J^{\pi}=3/2^-$
(paths (b) and (c) in Fig.~\ref{fig:levels}). Under these assumptions,
the effect on the capture yield would be -12\% for path (b) and \mbox{9\%} 
for path (c). However, mixtures of both decay paths would partly compensate 
the correction for angular distribution effects. Adopting one standard 
deviation of the two extreme cases $\sigma_\theta^{5/2^+} \simeq
10$\% would, therefore, represent a rather conservative estimate of the 
corresponding uncertainty. Nevertheless, even such a relatively large 
uncertainty for the cross section of $J^{\pi}=5/2^+$ resonances would
have negligible consequences for the Maxwellian averaged cross
section because these resonances contribute very little to the total
capture cross section (see Sec.~\ref{sec:macs}).

\subsection{\label{sec:summary uncertainties} Summary of uncertainties}
With the WFs calculated via the Monte Carlo technique, the accuracy 
of the PHWT has been investigated in detail by the n\_TOF collaboration
\cite{abb04}. It has been shown that the capture yield can
be determined from the measured raw data with an accuracy better than
2\%. 

Other  sources of systematic uncertainty pertaining to this measurement 
are due to the energy dependence of the neutron flux ($\pm$2\%) and to 
the background due to in-beam $\gamma$-rays ($\pm$1\%). In the particular 
case of the ($n, \gamma$) cross section of $^{206}$Pb, the uncertainty 
introduced by the angular distribution of the capture $\gamma$-rays
has to be considered as well. This effect has been estimated to
contribute an uncertainty of $\pm 4\%$ for resonances with $J^\pi = 3/2^-$ 
and less than $\pm 10\%$ for resonances with $J^\pi = 3/2^+, 5/2^+$.

\section{\label{sec:results}Results}

A total of 61 capture levels were analyzed in the neutron energy range from 3~keV up to 570~keV using the R-matrix code SAMMY
\cite{lar06}. In the analysis, the orbital angular momenta $l$ and the
resonance spins $J$ were adopted from Ref.~\cite{mug06}. Some of the $l$ and $J$
parameters listed in Table~\ref{tab:RP} are tentative or arbitrary if
missing in Ref.~\cite{mug06}. We list all the parameters used in our analysis
so that the final values can be recalculated if necessary.
The capture yield $Y(E_\circ,\Gamma_n,\Gamma_\gamma)$ 
was parameterized with the Reich-Moore formalism, and a channel radius 
of 9.5~fm was used for all partial waves. This parameterized yield was 
fitted to the corrected experimental yield by variation of the capture 
width $\Gamma_\gamma$ and/or neutron width $\Gamma_n$,

\begin{equation}\label{eq:Y}
f^t \times Y^{exp} = B + Y(E_\circ,\Gamma_n,\Gamma_\gamma),
\end{equation}

\noindent
where $f^t$ is the global yield correction factor given in
Sec.~\ref{sec:threshold}.
The term $B$ describing the background was parameterized as an analytical
function of the neutron energy in the range between 1~eV and
30~keV. Beyond 30~keV, $B$ was best described by means of a numerical
function (pointwise) determined from the measurement of the $^{208}$Pb sample (see
Ref.~\cite{dom06c} for details). The uncertainties quoted for the energy of
each resonance are only the statistical errors obtained from the fits of the
capture data performed with SAMMY.

\begin{longtable*}{ccccccccc}
\caption{Resonance parameters derived from the R-matrix analysis of the
  $^{206}$Pb($n, \gamma$) data.}\label{tab:RP}\\
\hline
\hline
$E_{\circ}$ & $l$ & $J$ & $\Gamma_{\gamma}$ & $\Delta{\Gamma_{\gamma}}$ & $\Gamma_{n}$ & $\Delta{\Gamma_{n}}$ & $K_r ^a$ & $\Delta{K_r}$ \\
(eV)        &     &     & (meV)             & (\%)                      &  (meV)      &  (\%) & (meV) & (\%)\\
\hline
3357.93(0.04) &  1 & 3/2 &  78.1 & 3 &  235  &  &117 & 2 \\
10865.0(0.4) &  1 & 3/2 &  64.9 & 9 &  44.1  & 8 &52.5 & 6 \\
11296.0(0.5) &  (1) & (1/2) &  455 & &  44.6  & 7 &40.6 & 7 \\
14220.0(0.6) &  1 & (1/2) &  152 & 6 &  1560  &  &139 & 5 \\
16428.0(0.4) &  0 & 1/2 &  2268 & 9 &  936  & 5 &662 & 5 \\
19744.0(1.3) &  1 & (1/2) &  156 & 7 &  2581  &  &147 & 6 \\
19809.0(0.9) &  1 & (3/2) &  295 & &  71.6  & 8 &115 & 6 \\
21885.0(0.9) &  1 & 3/2 &  121 & 6 &  875  &  &212 & 5 \\
25112.0(0.9) &  1 & 3/2 &  438 & 9 &  326  & 8 &374 & 6 \\
25428(5) &  1 & 1/2 &  254 & 7 &  48901  &  &253 & 7 \\
36200(6) &  1 & 1/2 &  312 & 14 &  35700  &  &309 & 13 \\
37480.0(1.9) &  1 & (3/2) &  151 & 15 &  890  &  &258 & 13 \\
39028(2) &  1 & (1/2) &  346 & &  93.0  & 36 &73.3 & 28 \\
40647(2) &  1 & (1/2) &  163 & 23 &  884  &  &138 & 19 \\
42083.0(1.7) &  1 & (3/2) &  419 & 21 &  1419  & 91 &647 & 26 \\
47534(2) &  (1) & (1/2) &  184 & 34 &  1000  &  &155 & 29 \\
59233.0(0.2) &  (2) & (3/2) &  322 & 16 &  1000  &  &487 & 12 \\
63976(3) &  (2) & 5/2 &  151 & 17 &  1110  &  &400 & 15 \\
65990(10) &  0 & 1/2 &  1186 & 9 &  82200  &  &1169 & 9 \\
66590(6) &  1 & 3/2 &  198 & 19 &  9530  &  &387 & 19 \\
70352(7) &  1 & 1/2 &  163 & 34 &  10780  &  &161 & 34 \\
80388(4) &  2 & 3/2 &  1490 & 8 &  7005  &  &2457 & 6 \\
83699(6) &  (2) & (3/2) &  351 & 16 &  8000  &  &673 & 15 \\
88509(6) &  2 & 5/2 &  375 & 13 &  7996  &  &1076 & 12 \\
91740(4) &  (1) & (3/2) &  298 & 25 &  1000  &  &460 & 19 \\
92620(13) &  0 & 1/2 &  991 & 15 &  32000  &  &961 & 15 \\
93561(6) &  2 & 3/2 &  125 & 37 &  7001  &  &246 & 37 \\
94743(7) &  2 & (3/2) &  241 & 20 &  7000  &  &465 & 20 \\
101220(7) &  2 & (5/2) &  119 & 26 &  8000  &  &351 & 25 \\
114380(5) &  1 & (3/2) &  655 & 24 &  2500  &  &1037 & 19 \\
114602(6) &  2 & (5/2) &  366 & 19 &  5600  &  &1030 & 18 \\
118100(6) &  2 & (5/2) &  390 & 16 &  5100  &  &1087 & 15 \\
124753(47) &  1 & 3/2 &  2972 & 9 &  300000  &  &5886 & 9 \\
125312(7) &  2 & (3/2) &  2783 & 10 &  21005  &  &4915 & 9 \\
126138(38) &  (1) & (3/2) &  319 & 32 &  100000  &  &635 & 32 \\
140570(23) &  2 & 3/2 &  1387 & 11 &  103000  &  &2736 & 11 \\
145201(6) &  (2) & (3/2) &  518 & 30 &  3100  &  &888 & 26 \\
146419(24) &  0 & 1/2 &  6092 & 8 &  176000  &  &5888 & 8 \\
150880(7) &  (1) & (1/2) &  554 & 48 &  4400  &  &492 & 43 \\
151290(13) &  2 & 5/2 &  457 & 23 &  19000  &  &1340 & 22 \\
191217(48) &  (1) & (1/2) &  767 & 28 &  96977  &  &761 & 27 \\
196990(37) &  1 & 1/2 &  584 & 45 &  64000  &  &579 & 44 \\
198618(34) &  2 & 3/2 &  2730 & 10 &  132108  &  &5350 & 10 \\
274630(22) &  1 & (1/2) &  514 & 65 &  32000  &  &506 & 64 \\
276984(49) &  2 & 3/2 &  2481 & 13 &  112000  &  &4854 & 13 \\
313400(18) &  2 & (3/2) &  1020 & 32 &  22000  &  &1950 & 31 \\
314340(84) &  2 & 5/2 &  964 & 24 &  179000  &  &2875 & 24 \\
356098(22) &  2 & (5/2) &  676 & 35 &  31000  &  &1985 & 35 \\
357465(87) &  2 & 3/2 &  1998 & 24 &  455000  &  &3979 & 24 \\
406200(55) &  2 & 5/2 &  656 & 51 &  102000  &  &1955 & 51 \\
407200(41) &  2 & 3/2 &  2906 & 24 &  71000  &  &5583 & 23 \\
416370(127) &  2 & 5/2 &  2722 & 16 &  307000  &  &8096 & 16 \\
433340(32) &  2 & (5/2) &  4122 & 17 &  47000  &  &11368 & 16 \\
434604(37) &  (2) & (3/2) &  4695 & 23 &  58000  &  &8687 & 21 \\
443412(13) &  (2) & (5/2) &  2375 & 21 &  14000  &  &6092 & 18 \\
466320(49) &  (1) & (3/2) &  5413 & 15 &  90000  &  &10211 & 14 \\
469080(76) &  2 & 3/2 &  3222 & 19 &  161000  &  &6317 & 19 \\
471789(28) &  (3) & (5/2) &  792 & 36 &  41000  &  &2330 & 35 \\
476310(172) &  0 & 1/2 &  5252 & 18 &  374000  &  &5180 & 17 \\
510690(51) &  (2) & (3/2) &  3123 & 18 &  86000  &  &6026 & 18 \\
572245(181) &  2 & 5/2 &  3838 & 13 &  793194  &  &11460 & 13 \\
\hline
\end{longtable*}
$^a$ Capture kernel $K_r = g \Gamma_{\gamma} \Gamma_n/\Gamma$, with
  $g=J+1/2$.\\

\subsection{\label{sec:comp} Comparison to previous work}
\begin{figure}[h]
\includegraphics[width=0.48\textwidth]{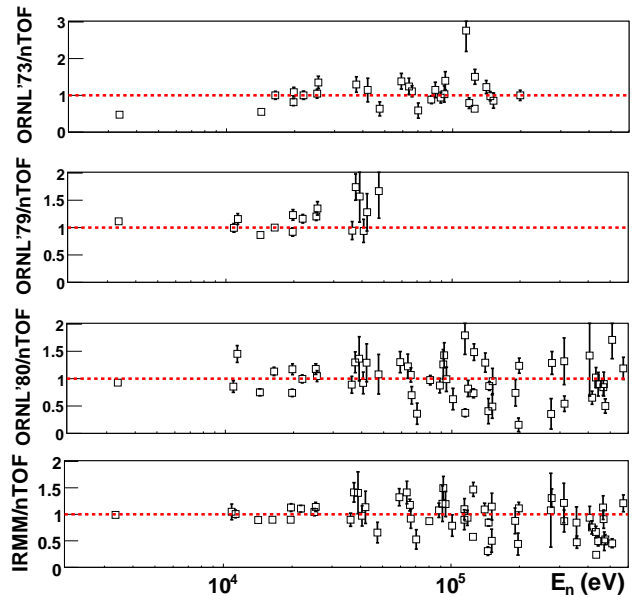}
\caption{\label{fig:ratios}(Color online) Ratio between the capture kernels reported in
  Ref.~\cite{all73} (top), Ref.~\cite{miz79} (2nd), Ref.~\cite{mus80} (3rd) and Ref.~\cite{bor07}
  (bottom) and the kernels determined here.}
\end{figure}

The radiative neutron capture cross section of $^{206}$Pb has been measured at
ORNL~\cite{mac64,all73,miz79}, at RPI~\cite{bar69} and at IRMM~\cite{bor07}. 
As representative examples of these measurements we consider in this section
two measurements made at ORELA~\cite{all73,miz79}, a more complete analysis~\cite{mus80} of the ORELA
capture data~\cite{all73} made in combination with transmission data~\cite{hor79} and the recent experiment
made at IRMM~\cite{bor07}.
In order to compare these four data sets with the present
results (Table~\ref{tab:RP}), the ratio of the capture kernels are shown in
Fig.~\ref{fig:ratios}. 

The values reported in Ref.~\cite{all73} show a relatively good agreement
with our results, except for the first two resonances at 3.3~keV and
14.25~keV, which are lower by $\sim$50\% (see
Fig.~\ref{fig:ratios}). However, these two resonances and the resonance at
16.428~keV are important because of their dominant contribution to the MACS
in the energy range between 5~keV and 20~keV. 
It is difficult to determine the source of discrepancy, thus no correlation has been found between the
discrepancies and the spins of the resonances. The latter could probably help to 
determine if there is any effect related to the angular distribution of the
prompt capture $\gamma$-rays or to the WF used in the previous measurement.

In the second measurement at ORELA~\cite{miz79} the discrepancies versus our present
results are smaller (see Fig.~\ref{fig:ratios}), but the capture kernels are
systematically larger, on average 20$\pm$5\% higher. This could probably reflect
that the WF used in Ref.~\cite{miz79} is overweighting the relatively hard
pulse height spectrum of $^{207}$Pb. Indeed, similar discrepancies have been
found in the past for $^{56}$Fe~\cite{mac87}, where the pulse height spectrum
is also considerably harder than that of the $^{197}$Au sample used for yield
normalization.

The posterior analysis~\cite{mus80} of the ORELA capture data~\cite{all73} in
combination with transmission~\cite{hor79} shows, on average, better agreement with the
capture areas reported here (see Fig.~\ref{fig:ratios}).

Finally, the results reported in the measurement at IRMM~\cite{bor07} show
the best agreement with the capture kernels of n\_TOF (see
Fig.~\ref{fig:ratios}). At E$_n \leq 40$~keV both
measurements agree within a few percent. At higher energy the fluctuations
are larger, but the agreement is still good within the quoted error bars.

\begin{figure}[h]
\includegraphics[width=0.45\textwidth]{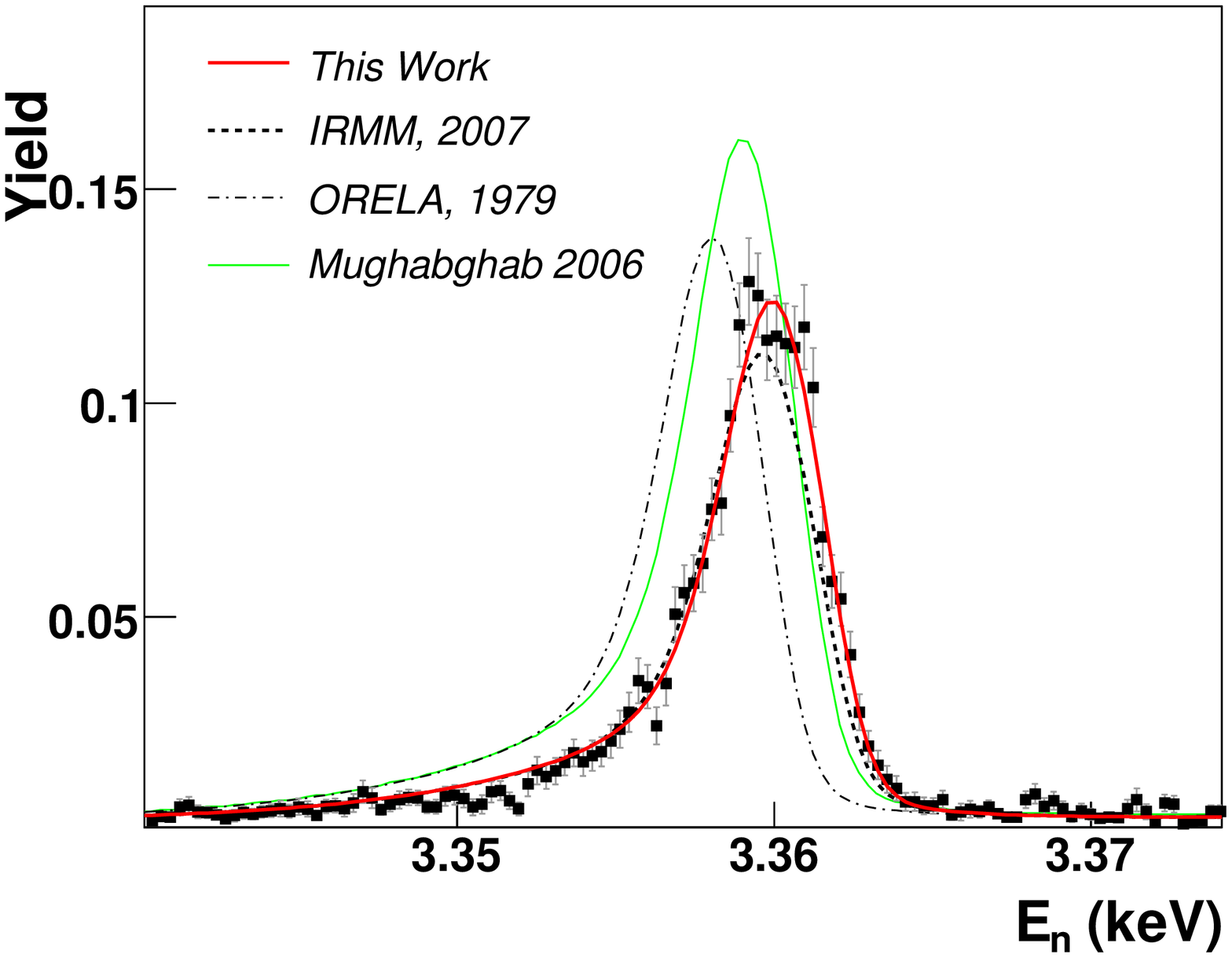}
\includegraphics[width=0.45\textwidth]{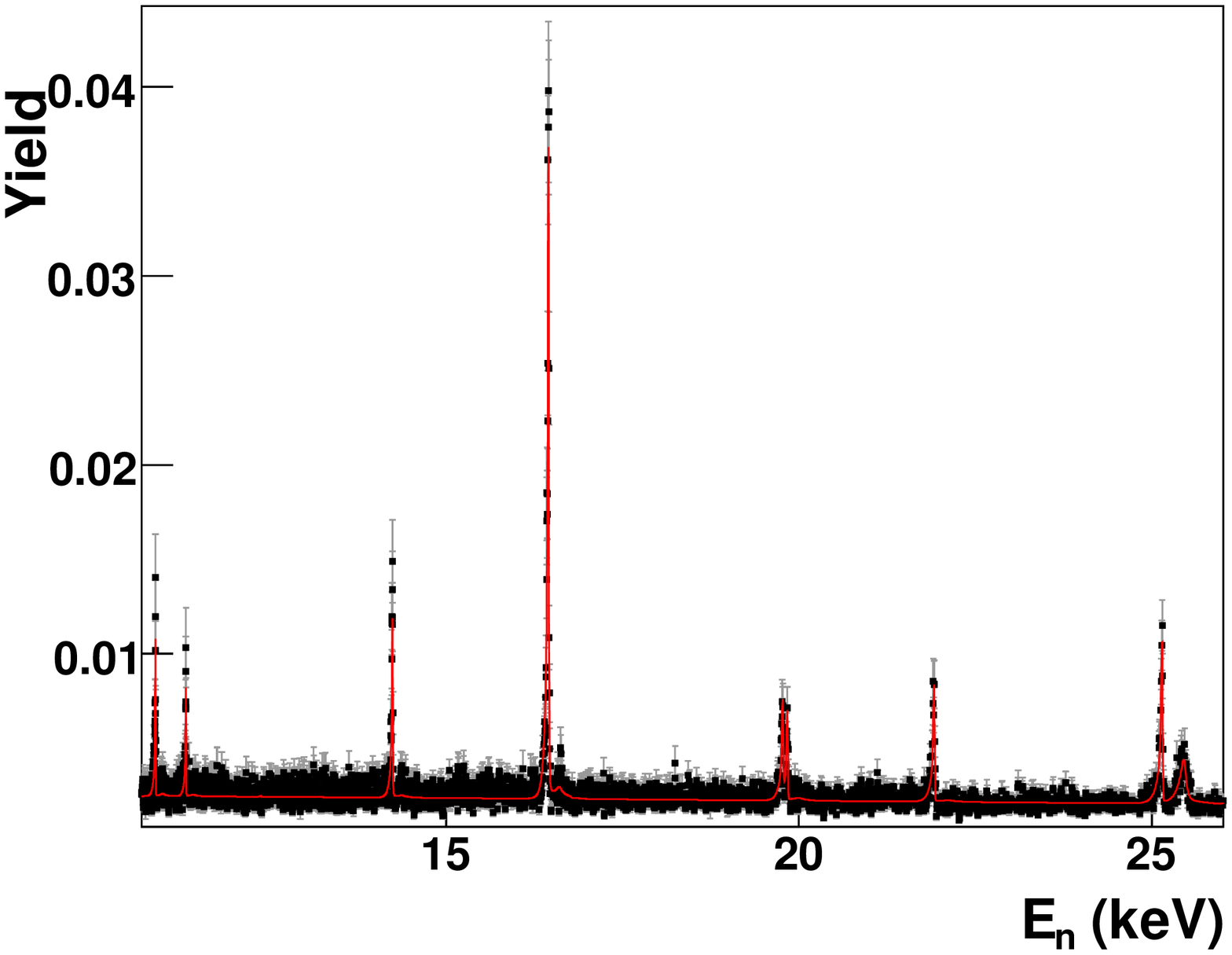}
\caption{\label{fig:resos}(Color online) (Top figure) The bold red line represents an R-matrix fit to our
  experimental capture yield starting from the initial parameters (solid green
  line) in Ref.~\cite{mug06}. The dashed and dott-dashed curves correspond to
  the capture yields determined in Ref.~\cite{bor07} and Ref.~\cite{miz79},
  respectively. (Bottom figure) The fitted capture yield in
  the 10-30~keV energy range (thin red line).}
\end{figure}

As an illustrative example, the capture yield measured
at n\_TOF for the first resonance at 3.3~keV is compared in the top panel of Fig.~\ref{fig:resos} versus the yield
calculated from the resonance parameters reported in
Refs.~\cite{bor07,miz79,mug06}. Obviously, the
IRMM and n\_TOF results show good agreement in both the capture area and
the resonance energy. 

\subsection{\label{sec:macs} Maxwellian averaged capture cross section}

The Maxwellian averaged cross section (MACS) was determined using the SAMMY
code in the range of thermal energies relevant for stellar nucleosynthesis, i.e. from $kT$=5~keV up to 
$kT$=50~keV. As discussed in the previous section, our results agree best with the values reported in
Ref.~\cite{bor07}. The latter data set seems also to be the most complete in
terms of number of analyzed resonances, with about 283 levels. Therefore our results were
complemented with resonances from Ref.~\cite{bor07} in order to avoid any
discrepancy due to resonances missing in Table~\ref{tab:RP}. The contribution
of these supplementary resonances to the MACS is $<0.1$\% at $kT=5$~keV and 6\% at
$kT=25$~keV. The fact that this correction starts to be significant towards 
$kT \gtrsim 25$~keV is not relevant for the study of the nucleosynthesis of
$^{206}$Pb. Indeed, as it is discussed below in Sec.~\ref{sec:implications},
$^{206}$Pb is mostly synthesized between the He-shell flashes of the
asymptotic giant branch stars. These intervals between pulses provide about
95\% of the neutron exposure via the $^{13}$C($\alpha, n$)$^{16}$O reaction, which operates
at a thermal energy of $kT = 8$~keV. At this stellar temperature less than
0.5\% of the MACS is due to the supplemented resonances.



\begin{figure}[h]
\includegraphics[width=0.45\textwidth]{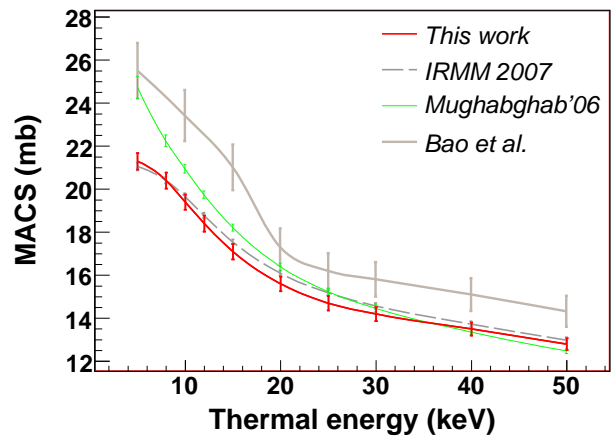}
\caption{\label{fig:macs} (Color online) Maxwellian averaged ($n, \gamma$) cross sections 
  for $^{206}$Pb from the resonance parameters of this work (bold red) compared
  to the IRMM measurement~\cite{bor07} (dashed), to the recommended data of
  Ref.~\cite{bao00} (grey) and to the compiled data of Ref.~\cite{mug06}
  (solid green).}
\end{figure}

The uncertainties shown in Fig.~\ref{fig:macs} are only statistical.
The systematic uncertainties of the MACS quoted in Table~\ref{tab:macs} 
include all contributions discussed in Sec.~\ref{sec:summary 
uncertainties}. 

\begin{table}[htbp]
\caption{\label{tab:macs} Maxwellian averaged cross section for
  $^{206}$Pb.}
\begin{ruledtabular}
\begin{tabular}{cccc}
Thermal energy $kT$ & MACS & $\sigma_{stat}$ & $\sigma_{sys}$\\
(keV)            & (mbarn) & (\%)           &(\%)\\
\hline
      5 &   21.3 &   1.8  & 5 \\
      8 &   20.4 &   1.8  & 3 \\
     10 &   19.4 &   1.9  & 3 \\
     12 &   18.4 &   2.0  & 3 \\
     15 &   17.1 &   2.1  & 3 \\
     20 &   15.6 &   2.2  & 3 \\
     25 &   14.7 &   2.3  & 3 \\
     30 &   14.2 &   2.3 &  4 \\ 
     40 &   13.5 &   2.2 &  4 \\ 
     50 &   12.8 &   2.1 &  4 \\
\end{tabular}
\end{ruledtabular}
\end{table}

Assuming systematic uncertainties of 4\% and 10\% 
for $3/2^-$ and $5/2^-$ resonances, respectively, the final
uncertainties are completely dominated by the 4\% uncertainty of 
the $3/2^-$ resonances. A change of 10\% 
in the cross section of the fewer $5/2^+$ resonances has a negligible 
influence on the MACS at $kT = 5$~keV, it contributes only 0.5\% at
$kT = 25$~keV and increases linearly up to 1\% at $kT = 50$~keV. 
An effect of 10\% in the capture yield of the $3/2^+$ resonances makes only a
1\% difference in the MACS at $kT = 25$~keV and it becomes also negligible
towards lower stellar temperatures.
The 3\% systematic uncertainty of the experimental method 
itself originates from the PHWT, the neutron flux shape,
and the use of the saturated resonance technique.

In summary, the MACS of $^{206}$Pb can now be given with total
uncertainties of 5\% and 4\% at the stellar temperatures corresponding to 5~keV
and 25~keV thermal energies, respectively. This improvement with respect to the previously
recommended values of Ref.~\cite{bao00} becomes particularly important for
determining the $s$-process contribution to the production of lead and bismuth in the Galaxy.

\section{\label{sec:implications}The $s$-process abundance of $^{206}{\rm \bf Pb}$ as a
constraint for the U/Th clock}

The $s$-process production of $^{206}$Pb takes place in low mass 
asymptotic giant branch (\textsc{agb}) stars of low metallicity
\cite{tra01}, where about 95\% of the neutron exposure is provided by 
the $^{13}$C($\alpha, n$)$^{16}$O reaction at a thermal energy 
of $kT \approx 8$~keV. At this stellar temperature the present 
MACS is about 20\% lower and two times more accurate (see Fig.~\ref{fig:macs}) than the values from
Ref.~\cite{bao00}, which have been commonly used so far for stellar nucleosynthesis
calculations. The additional neutron irradiation provided by the
$^{22}$Ne($\alpha, n$)$^{25}$Mg reaction at the higher thermal energy of
$kT=23$~keV during the He shell flash is rather weak.

With the new MACS the $s$-process abundance of $^{206}$Pb has been 
re-determined more accurately. A model calculation was carried out for
thermally pulsing \textsc{agb} stars of 1.5 and 3 $M_{\odot}$ and 
a metallicity of [Fe/H] = $-$0.3. The abundance of $^{206}$Pb is 
well described by the average of the two stellar models, which
represent the so-called main component~\cite{arl99}. Since the
contribution of $^{206}$Pb by the strong component is only 2\%,
the main component can be used to approximate the effective production 
of $^{206}$Pb during Galactic chemical evolution (GCE)~\cite{GAB98,tra99,tra01}. 
This approach yields an $s$-process abundance of $^{206}$Pb, which 
represents 70(6)\% of the solar abundance value $N_\odot^{206} = 0.601(47)/10^6$Si~\cite{lod03}. The same
calculation made with the older MACS recommended by Bao et al.~\cite{bao00}
yields 64\%. The uncertainty on the calculated $s$-process abundance is
mostly due to the uncertainty on the solar abundance of lead
(7.8\%)~\cite{and89b}. The contribution from the uncertainty on the MACS at
8~keV is less than 2\%. Finally, the contribution from the $s$-process model
is $\pm$3\%. The latter corresponds to the mean root square deviation 
between observed and calculated abundances for $s$-process only
isotopes~\cite{arl99}. This uncertainty is justified for $^{206}$Pb because
its nucleosynthesis is dominated by the main component and it is only
marginally affected ($\sim$2\%) by the strong component~\cite{GAB98,arl99,tra99,tra01}. 
Furthermore, because of the much lower cross sections
of $^{208}$Pb and $^{209}$Bi, the synthesis of $^{206}$Pb remains practically
unaffected by the $\alpha$-recycling after $^{209}$Bi~\cite{rat04}. This
lends further confindence that the production of $^{206}$Pb, and hence its 
uncertainty, follows the same trend as the main $s$-process component.

\begin{figure}[h]
\includegraphics[width=0.45\textwidth]{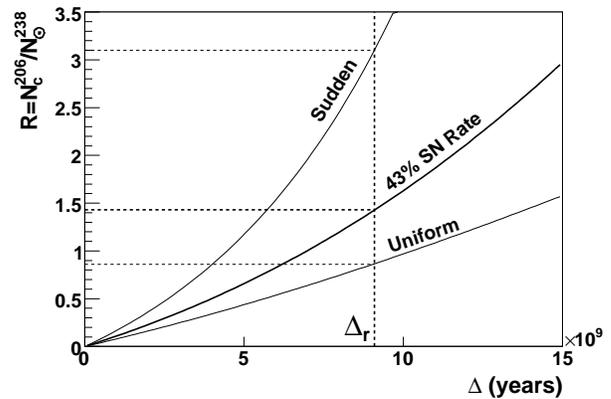}
\caption{\label{fig:Nc206} Estimate of the radiogenic component of $^{206}$Pb
  using the Fowler's model with different nucleosynthetic assumptions (see
  labels in curves) and the $r$-process age $\Delta_r = t_U - 4.6$~Gyr (vertical dashed line) derived from the age of the
  Universe $t_U$~\cite{ben03}.}
\end{figure}

In order to estimate a constraint for the $r$-process abundance of $^{206}$Pb
one needs to take into account its radiogenic contribution, $N_c^{206}$, due to
the decay of $^{238}$U. As it is shown in the following, this component is
relatively small but can not be neglected. Based on the schematic model of
Fowler, which assumes an exponential decrease of the $r$-process yield during
GCE~\cite{cla64} (supernova rate $\Lambda = (0.43t)^{-1}$~Gyr$^{-1}$) and using the current best estimates for the age of the
Universe ($t_U = 13.7 \pm 0.2$~Gyr)~\cite{ben03}, one obtains $N_c^{206} =
0.027(2)/10^6$Si (see Fig.~\ref{fig:Nc206} and Table~\ref{tab:N}). This number, combined with our result for $N^{206}_s$,
yields an $r$-process residual,

\begin{equation}\label{eq:n206}
N^{206}_{r} = N^{206}_{\odot} - N^{206}_s - N^{206}_c = 0.153\pm0.063.
\end{equation}

The uncertainty in this result includes contributions of 8.4\% from $N^{206}_c$
(corresponding to the uncertainty on the solar abundance of $^{238}$U~\cite{and89b}), 7.8\% from the total solar
abundance of $^{206}$Pb, $N^{206}_{\odot}$~\cite{lod03,and89b}, and 8.6\% from the
determination of $N^{206}_{s}$ as discussed above. This means that, apart from
the uncertainties related with the simplified assumptions in the GCE model of Fowler, the
$r$-process abundance can be reliably constrained between 16\% and 36\% of
the solar $^{206}$Pb. 

\begin{table}
\caption{\label{tab:N} Radiogenic abundance of $^{206}$Pb,
  $N_c^{206}$ (Si=10$^6$), derived from the model of Fowler and the age of the Universe
  (see Fig.~\ref{fig:Nc206}). $r$-Process residuals obtained via eq.~\ref{eq:n206}.}
\begin{ruledtabular}
\begin{tabular}{lccc}
GCE & $N^{206}_c = R N^{238}_\odot$  &\multicolumn{2}{c}{$ N^{206}_r =  N^{206}_{\odot} - N^{206}_s - N^{206}_c$}\\
(Fig.~\ref{fig:Nc206}) & $10^6$Si & $10^6$Si & $N^{206}_r/N^{206}_\odot$(\%)\\ 
\hline
43\% SN rate     & 0.027(2)   & 0.15(6) & 26(10) \\
Sudden           & 0.058(5)   & 0.12(6) & 20(10) \\
Uniform          & 0.0161(14) & 0.16(6) & 27(10) \\
\end{tabular}
\end{ruledtabular}
\end{table}

The $r$-process residuals derived here are consistent with $r$-process
model calculations available in the literature, i.e., $N^{206}_r =
26.6$\%~\cite{cow99}. More recent calculations yield $N^{206}_r$ values between 27\%
and 35\%~\cite{kra04}. 
One can also derive hard limits for the $r$-process abundance, considering the two
extreme cases of sudden nucleosynthesis ($\Lambda \rightarrow \infty$) and
uniform nucleosynthesis ($\Lambda \rightarrow 0$). This
yields constraints between 10\% and 37\% of solar $^{206}$Pb (see
Table~\ref{tab:N}).

The situation is rather different for the corresponding $^{207}$Pb/$^{235}$U 
ratio, which has been investigated as a potential clock in the
past~\cite{bee85}. 
In this case, the $s$-process abundance of $^{207}$Pb was recently 
determined to be $N^{207}_s = 77(8)$\%~\cite{dom06b}. A similar calculation
to that shown in Fig.~\ref{fig:Nc206} gives $N_c^{207} = 0.150(13)$ (see Table~\ref{tab:N207}). The
latter value reflects the large relative radiogenic abundance of $^{207}$Pb, $N_c^{207}/N_\odot^{207} =
22$\%, due to the much shorter half-life of $^{235}$U. 
From the total solar abundance of $^{207}$Pb~\cite{lod03} and the $N_s^{207}$
and  $N_c^{207}$ values quoted above, the $r$-process residual becomes $N^{207}_r = 0.003 \pm 0.073$,
which means that $N^{207}_r$ can not be larger than 11\% of the $^{207}$Pb
abundance in the solar system, $N_\odot^{207} = 0.665(52)$~\cite{lod03} (Table~\ref{tab:N207}). 

\begin{table}
\caption{\label{tab:N207} Radiogenic abundance of $^{207}$Pb,
  $N_c^{207}$ (Si=10$^6$), derived from the model of Fowler and the age of
  the Universe. $r$-Process residuals obtained via eq.~\ref{eq:n206}.}
\begin{ruledtabular}
\begin{tabular}{lccc}
GCE & $N^{207}_c = R N^{235}_\odot$  &\multicolumn{2}{c}{$ N^{207}_r =  N^{207}_{\odot} - N^{207}_s - N^{207}_c$}\\
 & $10^6$Si & $10^6$Si & $N^{207}_r/N^{207}_\odot$(\%)\\ 
\hline
43\% SN rate     & 0.150(13)   &  0.003(73)  & 0(11) \\
90\% SN rate     & 0.08(7)   &  0.073(72)  & 11(11) \\
Uniform          & 0.047(4)    &  0.106(72)  & 16(11) \\
\end{tabular}
\end{ruledtabular}
\end{table}

This result is in contrast with $r$-process
model calculations, which yield values between 22.7\% and
25.3\%, with a relative uncertainty of 15-20\%~\cite{cow99,kra04}. 
The $s$-process abundances of $^{206,207}$Pb are rather reliable
and not very sensitive to details of the stellar models~\cite{rat04,dom05}. 
Therefore, this discrepancy indicates that $r$-process abundances 
might have been overestimated, possibly because the odd-even
effect is not properly reproduced by the ETFSI-Q mass model implemented 
in the $r$-process calculations~\cite{cow99,kra04}. Indeed, one needs to
increase the supernova rate in the standard Fowler model from 43\% up to 90\% ($\Lambda =
(0.90t)^{-1}$~Gyr$^{-1}$) in order to achieve
agreement between these $r$-process constraints and the latter $r$-process
calculations~\cite{cow99,kra04}. Obviously the less realistic uniform
scenario would also provide agreement with the abundances from these
r-process models (see Table~\ref{tab:N207}).

However the situation has been improved recently after more detailed $r$-process model
calculations~\cite{kra07} predicted a new $N_r^{207}$ value, which is 35\% lower
than the previous one of Ref.~\cite{cow99}. This yields $N^{207}_r/N^{207}_\odot = 16.8$\%,
which is substantially closer (considering an uncertainty of 20\%) to the upper limit of 11\% derived here. In
this case a good agreement would be found for a more reasonable increase of
the supernova rate to 55\% in the Fowler model.

These constraints for the $r$-process abundances of $^{206,207}$Pb
become relevant for the validation of $r$-process model calculations and
hence, for the reliable interpretation of actinide abundances
observed in UMP stars and their use as cosmochronometers.

The $s$-process aspects will be more rigorously investigated
in a comprehensive study of the Pb/Bi region~\cite{bis06}, where the role of
stellar modeling and GCE will be discussed with a complete set of new cross
sections for the involved isotopes, including the present data for
$^{206}$Pb, and recent results for $^{204}$Pb~\cite{dom06c},
$^{207}$Pb~\cite{dom06b} and $^{209}$Bi~\cite{dom06}.

\section{\label{sec:summary} Summary}
The neutron capture cross section of $^{206}$Pb as a function of the neutron
energy has been measured with high resolution at the CERN n\_TOF installation using 
two C$_6$D$_6$ detectors. Capture widths and/or radiative kernels could 
be determined for 131 resonances in the neutron energy interval from 
3~keV up to 620~keV. Systematic uncertainties of 3\%, 5\%, and 
$\lesssim$10\% were obtained for resonances with spin-parities of 
1/2$^{\pm}$, 3/2$^-$ and 5/2$^+$, respectively. The Maxwellian averaged 
cross sections were found to be significantly smaller by 10\% to 20\% 
compared to values reported earlier~\cite{bao00}, resulting in a 
correspondingly enhanced $s$-process production of $^{206}$Pb. First 
calculations with a standard \textsc{agb} model yield an $s$-process 
component of 70(6)\% for the $^{206}$Pb abundance. 
Combined with an estimate of the radiogenic production of $^{206}$Pb, the
$r$-process abundance is constrained between 16\% and 36\% of the solar
$^{206}$Pb abundance, well in agreement with $r$-process model calculations
reported in the literature~\cite{cow99,kra04}. A similar analysis for $^{207}$Pb shows
agreement only with most recent $r$-process model calculations~\cite{kra07}.

\bibliography{bibliographyv2}

\end{document}